\documentstyle{l-aa}

\begin{document}

\thesaurus{06         % A&A Section 6: Form. struct. and evolut. of stars
              (03.11.1;  % Cosmogony,
               16.06.1;  % Planets and satellites: general,
               19.06.1;  % Solar system: general,
               19.37.1;  % Stars: formation of,
               19.53.1;  % Stars: oscillations of,
               19.63.1)} % Stars: structure of.
\title{A new investigation on the Antlia Dwarf Galaxy\thanks{Based on 
observations collected with VLT-UT1 telescope of ESO in Paranal, during 
the Science Verification Program}}
%   \subtitle{}

\author{A. M. Piersimoni,\inst{1} G. Bono,\inst{2} M. Castellani,\inst{2} 
G. Marconi,\inst{2} S. Cassisi,\inst{1} R. Buonanno,\inst{2} 
and M. Nonino\inst{3,4}}

\institute{$^1$Osservatorio Astronomico di Collurania, via M. Maggini, 
64100 Teramo, Italy\\
$^2$Osservatorio Astronomico di Roma, via Frascati 33, 
00040 Monte Porzio Catone, Italy\\
$^3$European Southern Observatory, D-85748 Garching bei M\"unchen, Germany\\
$^4$Osservatorio Astronomico di Trieste, Via G.B. Tiepolo 11, 
40131 Trieste, Italy} 
          
\offprints{A.M. Piersimoni}

%\date{Received September 15, 1999; accepted March 16, 1999}

\maketitle
\markboth{Piersimoni et al.: The Antlia Dwarf Galaxy}
\begin{{\bf Abstract}}

We present deep $(I,V-I)$ and $(I,B-I)$ color-magnitude diagrams (CMDs) 
of the Antlia dwarf galaxy, based on Science Verification (SV) data 
collected with the FORS I camera on the ESO Very Large Telescope (VLT).
The CMDs present two key features: a well-defined Red Giant Branch (RGB), 
and a sample of bright blue stars, belonging to a young stellar component.
The comparison between theory and observations confirms that this sample
of bright stars is consistent with the occurrence of a star formation 
episode $\approx0.1$ Gyr ago. In agreement with previous investigations
(Sarajedini et al. 1997; Aparicio et al. 1997), we also find that this 
young stellar population is more centrally concentrated than the old one.

By adopting the new calibration of the Tip of the RGB (TRGB) provided 
by Salaris \& Cassisi (1998) we estimated that the Antlia distance 
modulus is $(m-M)_0=25.89\pm0.10$ mag, and therefore a distance 
$D=1.51\pm 0.07$ Mpc. This distance determination is  $\approx13\%$ 
larger than the values suggested in previous investigations. 
By adopting the calibration of the RGB $(V-I)$ color index as a
function of the metallicity, we estimated that the mean metallicity of 
Antlia stellar population is of the order of $[Fe/H]\approx-1.3$.
This metallicity estimate is at least 0.3 dex more metal-rich than 
similar evaluations available in literature. The disagreement 
both in the distance and in the metallicity determinations are 
mainly due to difference in the calibration of the TRGB method
and in the $(V-I)$ color index vs. metallicity relation.  
 
The differential RGB luminosity function shows an excess in the 
observed counts -at the $2\sigma$ level- when compared with 
theoretical predictions. Plain arguments on the dependence of the 
TRGB luminosity on stellar age suggest that this discrepancy might 
be due to a stellar component with an age approximately equal to 
0.7 Gyr. 

%\end{abstract} 

%%%%%%%%%%%%%%%%%%%%%%%%%%%%%%%%%%%%%%%%%%%%%%%%%%%%%%%%%%%%%%%%%%%%%%%%%%%%%%%
\section{Introduction}

The global properties of Local Group (LG) galaxies play a key role for 
understanding both galaxy formation and evolution. Among them the dwarf 
galaxies (DGs) are particularly interesting since they are a unique 
laboratory to address several open questions on low-luminosity galaxies.  
In particular, the global properties of nearby DGs are crucial to understand 
the relationship, if any, between different morphological-types 
(Minniti \& Zijlstra 1996) as well as to estimate how fundamental 
parameters such as the dark matter content, the chemical composition, 
and the star-formation history depend on the global luminosity, and 
in turn on the total mass of the DGs (Grebel 1998; Mateo 1998; 
van den Bergh 1999a). Moreover, since the LG contains both fairly 
isolated galaxies and dwarfs in subgroups, it allows us to investigate 
the environmental effects on the galactic evolution.
As a consequence a detailed analysis of the evolutionary properties of 
their stellar component(s) is a fundamental step to shed new light not 
only on the formation and the interaction of the Galaxy and of M31 with 
their satellites but also for understanding distant, unresolvable stellar
systems such as the very low surface brightness galaxies 
(Whiting, Irwin, \& Hau 1997, hereinafter WIH; Grebel 1998).
The observational scenario on nearby DGs was further enriched by the 
evidence that the Antlia-Sextans clustering may be the nearest group 
of galaxies not bound to the LG (van den Bergh 1999b, hereinafter VDB). 

Ground based data provided several deep and accurate CMDs for
not-too-distant DGs (Smecker-Hane et al. 1994; Marconi et al. 1998),
while data collected by the Hubble Space Telescope (HST) were 
crucial for assessing the stellar content of distant LG galaxies
(see e.g., Mighell \& Rich 1996; Buonanno et al. 1999; 
Caputo et al. 1999; Gallart et al. 1999, and references therein).
Even though HST data provided high-quality CMDs for a large sample 
of DGs in the LG, in the near future the use 8-10m telescopes 
can substantially improve our knowledge of their global properties.
In fact, the large collecting area and the evidence that 
DGs are only marginally affected by crowding problems even in the 
innermost regions make this class of instruments particularly  
useful for investigating the stellar content in the LG galaxies.
In this paper, we present the results of an investigation on the 
Antlia dwarf galaxy based on B,V,I data collected 
with FORS I on VLT during the SV program. These photometric data 
are a plain evidence of the VLT capability to investigate DGs 
in the LG (but see also Tolstoy 1999).

The layout of this paper is the following: in the next section we briefly 
present the observations and describe the procedures adopted for the 
reduction and the calibration of data. In \S 3 we discuss the 
main features of the CMDs, together with the distance and metallicity 
estimates, while the Antlia-Sextans grouping is addressed in \S 4. 
Finally, a brief summary and the conclusions are outlined in \S 5.

%%%%%%%%%%%%%%%%%%%%%%%%%%%%%%%%%%%%%%%%%%%%%%%%%%%%%%%%%%%%%%%%%%%%%%%%%%%%%%
\section{Observations and data reduction}

The data for Antlia have been requested and retrieved electronically
from the ESO archive in Garching.
The galaxy was observed through the standard Bessell B,V,I filters 
during the VLT-UT1 SV Program in January 1999 using the FORS I camera 
which covers a $6.8\times6.8$ arcmin field of view at 0.02 arcsec per
pixel resolution. The seeing was excellent (0.45 - 0.75 arcsec). 
We used the standard reduction procedure reported on the Data 
Reduction Notes listed in the VLT web pages and the Daophot II 
package (Stetson, Davis, \& Crabtree 1990, and references therein). 
To improve the detection limit we coadded all the frames taken with 
the same filter and then we selected the I coadded frame (5400 sec) 
to create a master catalogue of stellar objects. The stars identified 
in this search were used as a template to fit both the B and the V 
coadded frames.

By adopting a SHARP parameter of $\pm0.5$, we detected in the coadded 
frames 3711 (B), 2958 (V), and 4583 (I) stars down to $B\approx27.0$ mag, 
$V\approx25.7$ mag, and $I\approx25.5$ mag respectively.  
The calibration was derived by using a standard field observed during the 
same night and by adopting the average extinction coefficients. 
Completeness tests were performed by randomly adding in each 
magnitude bin (0.2 mag) 15-20\% of the original number of stars 
to the coadded frames. 
Only stars that were detected in the same position and within 
a magnitude bin of $\pm0.1$ mag were considered as recovered. 
The simulations we performed suggest that a completeness of the 
order of 50\%  was reached at $I\approx 24.0$ mag and 
$B\approx 25.1$ mag respectively.

%%%%%%%%%%%%%%%%%%%%%%%%%%%%%%%%%%%%%%%%%%%%%%%%%%%%%%%%%%%%%%%%%%%%%%%%%%%%%%
\section{The Color-Magnitude diagram}

The Antlia dwarf galaxy was originally noted by Corwin, de Vaucouleurs
\& de Vaucouleurs (1985) and by both Feitzinger \& Galinski (1985) and
Arp \& Madore (1987) who also suggested that this stellar system could 
be a nearby galaxy. This finding was subsequently confirmed by  
Fouqu\'e et al. (1990) who found, in a detailed $HI$ survey of 
southern late-type galaxies, that Antlia has a small radial 
velocity ($V_r=361\pm2 km s^{-1}$).
However, a firm identification of Antlia was only recently provided 
by WIH in a systematic search for VLSB galaxies in 894 ESO-SRC IIIaJ 
plates covering the entire southern sky, who also suggested that this
galaxy is probably gravitationally bound to the dwarf irregular galaxy 
NGC3109. 
After its rediscover the global properties of this galaxy were 
investigated by Aparicio et al. (1997, hereinafter ADGMD) and by 
Sarajedini, Claver \& Ostheimer (1997, hereinafter SCO). These 
investigations brought out the following characteristics: 
a) low-mass, metal-poor stars are the main stellar component of 
the galaxy; b) evidence of an age gradient within the galaxy with 
the young stellar component located close to the center; 
c) no evidence of an ongoing star formation process.

A peculiar feature of Antlia is the amount of gas still present in 
this galaxy, and indeed WIH have inferred a total {\it H I} mass 
of $8\times10^5 M_\odot$. Following the classification suggested
by Da Costa (1997), which is based on the ratio between total mass 
of gas and B integrated luminosity, Antlia should be classified as 
a (dusty) dwarf irregular (dIrr) rather than as a dwarf spheroidal 
(dSph) galaxy, since according to SCO it should also contain 
interstellar dust. Oddly enough, it also shows a smooth elliptical 
morphology and a very low stellar concentration in the innermost regions. 
These features are quite similar to other isolated dSph galaxies in the LG 
such as the Tucana dwarf. However, Tucana does not show a significant 
amount of gas and therefore Mateo (1998) classified Antlia as 
a {\em transitional galaxy} (dIrr/dSph) together with LGS3, Phoenix, 
DDO210, and Pegasus.

Panels a) and b) of Figure 1 show the CMD of Antlia in the $(I, V-I)$ 
and in the $(I, B-I)$ CMD respectively. In these diagrams were only 
plotted stars ($\approx1700$) located within 2.5 arcmin from the 
center of the galaxy. We selected the stars whose centroids in 
B, V, and I frames were matched in one pixel (0.2 arcsec) and which 
satisfy tight constraints on photometric accuracy, namely  
$\sigma_I\le 0.2$, $\sigma_{B-I}\le 0.3$, 
$\sigma_{V-I} \le 0.3$, and SHARP parameter $\le \pm0.5$. 

Figure 1 discloses several interesting features. It is noteworthy 
the well-populated RGB, extending from $I\approx24$ to $I\approx21.5$ mag, 
and the sizable number of stars brighter than the TRGB. 
SCO identified the latter objects as a not-negligible population 
of Asymptotic Giant Branch stars, progeny of an intermediate-age 
population. On the basis of stellar counts in nearby fields (ADGMD), 
and of the evidence that these bright stars are not particularly 
concentrated toward the center of the galaxy -they appear at distances 
larger than 1 arcmin- the possibility that they are actually 
foreground objects cannot be ruled out. 

%________________________________________________________________________ 
\begin{figure*}
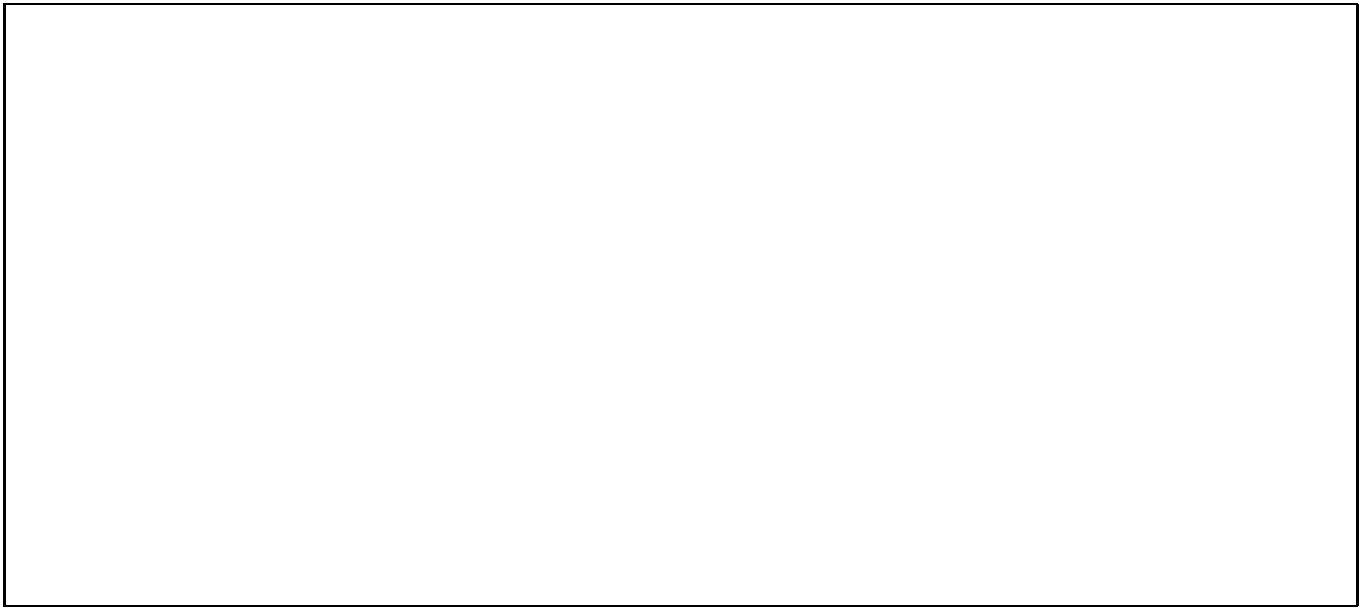

\picplace{8cm}
\caption{
{\it Panel a}): The $(I,V-I)$ CMD of Antlia. 
{\it Panel b}): Same as panel a) but for the $(I,B-I)$ CMD. 
The two redder RGB loci refers to stellar population with an age 
of 14 (solid line) and 0.8 Gyr (dashed line) respectively. The two 
isochrones plotted in the blue region refer to stellar ages 
of 150 (solid line) and 80 Myr (dashed line) respectively.  
Theoretical prescriptions were plotted in the observative plane 
by assuming a distance modulus of $(m-M)_0=25.89$ mag and a 
metallicity equal to $[M/H]= -1.3$.}  
\label{}
\end{figure*}
%______________________________________________________________________ 

By accounting for the RGB intrinsic width, SCO suggested that Antlia 
could contain a sizable amount of interstellar dust.
This suggestion was mainly based on the slope of the RGB upper portion 
which in their $(I, V-I)$ CMD mimic the slope of objects affected by 
increasing reddening. This feature seems not confirmed by present 
photometry. The $(I, B-I)$ CMD is particularly compelling since in 
this plane the slope of the reddening vector is steeper than in 
the $(I, V-I)$ CMD (see also ADGMD). Even though the origin of such 
a discrepancy cannot firmly established, we suggest that the photometric
error (the FWHM of SCO data is roughly a factor of two larger than 
in our data) and the contamination with foreground objects might 
have introduced a spurious trend (see \S 4.2 in SCO).   

The sample of stars located in the blue region of the CMDs 
-$(V-I)<0.7$, $(B-I)<1$- plotted in Figure 1 is quite interesting. 
This feature already found by SCO and by ADGMD suggest the presence 
of a young stellar population. Moreover, their radial distribution 
clearly shows that this sample is strongly concentrated in the 
innermost regions of the galaxy. Therefore we confirm the difference 
in the radial distribution  between the young and the old stellar 
component found by SCO and by ADGMD.  
The presence of this young stellar component together with the 
evidence of sizable amount of gas are the most clear indications
that Antlia should be classified as a {\em dIrr} rather than 
as {\em dSph} galaxy.

In order to supply an estimate of the age of the blue stars we plotted 
in the CMDs of Figure 1 evolutionary prescriptions (Cassisi 1999) for  
H and He-burning phases at two different stellar ages, namely 
$t=80$~Myr and $t=150$~Myr. At the same time, we also plotted 
the location of the RGB for two old stellar populations at 
$t\approx14Gyr$ and $t\approx0.8Gyr$ respectively.  
Theoretical predictions were transformed into the observational plane 
by adopting the bolometric corrections and the color-temperature 
relations provided by Green (1988). The adopted distance modulus, 
metallicity and reddening are discussed in the next section. 
The comparison between theory and observations 
clearly shows that the position of bright blue stars in Antlia is 
finely reproduced by an isochron with an age ranging from 100 
to 150~Myr. This age range implies that the TO masses of blue stars 
range from 3.5 to 4.5 $M_\odot$ and therefore that in this galaxy 
should be present classical Cepheids with periods of the order of 
few days.

%%%%%%%%%%%%%%%%%%%%%%%%%%%%%%%%%%%%%%%%%%%%%%%%%%%%%%%%%%%%%%%%%%%%%%%%%%%%%%%
\subsection{Distance and metallicity.}

When dealing with composite stellar population systems for which only the
bright end of the CMD is well sampled, as in the case of Antlia, the TRGB 
method turns out to be a valuable distance indicator. 
In fact, this standard candle  
works for all morphological types of galaxies as long as an old stellar 
population is present. Moreover the absolute I-Cousins magnitude of the 
TRGB presents a negligible dependence on metal content over a wide 
metallicity range, at least for $[M/H]<-0.5$. After the first 
semi-empirical calibration by Lee, Freedman \& Madore (1993, 
hereinafter LFM), Salaris \& Cassisi (1997, 1998, hereinafter SC97 
and SC98) provided a new theoretical calibration of the TRGB method, 
characterized by a magnitude shift of $\approx0.15$ -toward brighter 
magnitudes- in the absolute I magnitude of the tip. Note that Antlia 
distance estimates available in the literature are based on the TRGB 
calibration provided by LFM, while we here adopt the SC98 calibration. 

To estimate the apparent magnitude of the TRGB we use the differential 
luminosity function (LF) of RGB stars. 
We have not performed any correction since as clearly shown by ADGMD 
the stars located close to the tip are only marginally affected by 
foreground contamination. The top panel of Figure 2 shows the 
differential LF evaluated by adopting a magnitude bin of 0.20 mag. 
The error bars for each bin were estimated taking into account 
both the statistical fluctuations and the corrections for completeness. 
In order to obtain a robust determination of the RGB tip discontinuity,  
the LF was convolved with an edge-detecting Sobel filter 
$\rm [-1,0,+1]$. This function shows a sharp peak at 
$I_{TRGB}=21.7\pm0.10$ mag which marks the appearance of the 
RGB tip. The error was estimated on the basis of the adopted 
magnitude bin. This value is, within current observational uncertainties, 
in good agreement with the values obtained by SCO ($I_{TRGB}=21.63\pm0.05$) 
and ADGMD ($I_{TRGB}=21.64\pm0.04$). Thus supporting {\it a posteriori} 
the negligible effect of foreground contamination on the LF of RGB 
stars. Our determination is $\approx0.2$ mag fainter than the estimate 
provided by WIH, i.e. $I_{TRGB}=21.4\pm0.1$. This discrepancy was already 
noted by ADGMD who suggested that it could be due to crowding problems 
at the limiting magnitude in the WIH's photometry.  

%------------------------------------------------------------------------------ 
\begin{figure}[htbp]
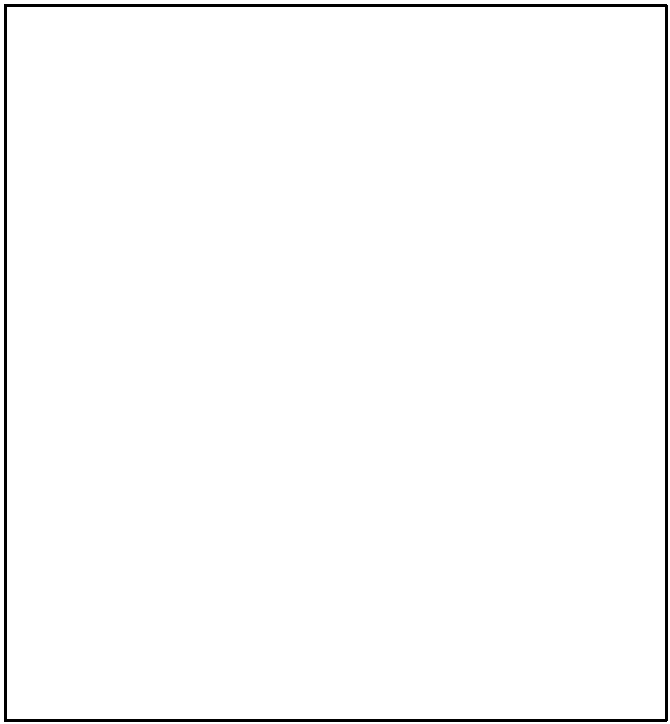

\picplace{9.5cm}
\caption{Logarithmic differential LF of RGB stars (solid line) in Antlia.  
The arrow marks the position of the bin where the TRGB discontinuity was 
detected. A theoretical LF (dashed line) for an age of 14 Gyr is also 
shown by adopting a distance modulus equal to $(m-M)_0=25.89$ mag. 
The theoretical LF was normalized to the observed one, at $I\approx21.9$ mag. 
{\em Bottom panel}: The cumulative LF of RGB stars. The dashed lines show 
the two different slopes of the LF.}
\label{}
\end{figure}
%------------------------------------------------------------------------------ 

By interpolating the maps of Burstein \& Heiles (1982) we estimated 
$E(B-V)=0.03\pm0.02$, which implies a foreground reddening of 
$E(V-I)=0.04\pm0.03$. 
By using the extinction relation provided by Cardelli et al. (1989), 
this reddening implies an extinction $A_I=0.04\pm0.03$ mag. 
As a consequence, the uncertainty on the I magnitude of the  RGB tip 
is dominated by photometric and completeness errors since the reddening 
correction for this band is significantly smaller 
($I_{TRGB,0}=21.66\pm0.10$ mag). This notwithstanding, our estimate of 
$I_{TRGB,0}$ is in good agreement with the values suggested by 
SCO ($I_{TRGB,0}=21.57$ mag) and by ADGMD ($I_{TRGB,0}=21.57\pm0.05$ mag).  

The TRGB method (see LFM, SC97 and SC98) is an iterative procedure which 
simultaneously gives both the distance and the mean metallicity of the 
old stellar population in the galaxy. The metallicity 
evaluations are based on the calibration of the dereddened color index 
$(V-I)$ of the RGB half a magnitude below the tip, $(V-I)_{-3.5,0}$, 
as a function of the metal content. 
By adopting the SC98 calibrations we find a distance modulus of 
$(m-M)_0=25.89\pm0.10$ mag, i.e. $D=1.51\pm 0.07$ Mpc and 
$(V-I)_{-3.5,0}$=1.36 mag, which in turn translates into a mean 
metallicity of $[M/H]\approx-1.3\pm0.15$. 
This distance modulus is roughly 13\% larger than the values 
found both by SCO ($(m-M)_0=25.62\pm0.12$ mag) and by ADGMD 
($(m-M)_0=25.6\pm0.1$ mag). As far as the mean metallicity is 
concerned, our estimate is $\approx0.3$ dex higher than the ADGMD 
evaluation ($[M/H]\approx-1.6\pm0.1$) and $\approx0.6$ dex higher 
than SCO determination ($[M/H]\approx-1.9\pm0.13$).  

As expected the disagreement is mainly due to the calibration of the 
TRGB method and of the RGB color index vs. metallicity provided 
by LFM and by SC98. A thorough discussion of the differences between 
these calibrations was already provided by SC98. However, 
the discrepancy with the metallicity evaluation provided by SCO 
is mainly due to the different approach adopted by these authors
for estimating the metallicity along the RGB. 
In passing we note that our metallicity estimate is in fair agreement 
with the metallicity obtained by comparing the RGB loci of Antlia 
directly with the RGB loci of galactic globular clusters provided 
by Da Costa \& Armandroff (1990).

%%%%%%%%%%%%%%%%%%%%%%%%%%%%%%%%%%%%%%%%%%%%%%%%%%%%%%%%%%%%%%%%%%%%%%%%%%%%%%%
\subsection{Some hints on RGB stars.}

The observed color width of the Antlia RGB at $I\approx22.1$ mag
-i.e. $\approx0.5$ mag below the RGB tip- is $\Delta(V-I)\approx0.25$ mag.
However, since at $I\approx22.1$ mag our mean photometric error is
$\sigma_{V-I}=0.05$ mag, it turns out that the intrinsic color width
of the RGB is roughly equal to 0.2 mag.
If we assume that this color dispersion is due to a spread in metallicity, 
the metal content of Antlia should be in the range $-1.8 <[M/H]< -1.0$. 
However, stellar models supply important information to figure out this 
problem. In fact, at fixed metal content an increase in age moves the 
RGB loci toward redder colors, while a decrease in age causes a decrease 
in the TRGB luminosity (see e.g. SC98; Caputo et al. 1999). 
Figure 1 shows the theoretical prescriptions at fixed metallicity 
-$[M/H]=-1.3$- for the RGB loci of an old (14~Gyr, solid line) and 
a young (0.8~Gyr, dashed line) stellar population. As a result, 
the RGB color dispersion in Antlia could be due to a mix of an old 
and of a young/intermediate-age stellar population. 

The LF shows a further interesting feature: for magnitudes dimmer than 
$I\approx22.3$ mag one can notice a significant increase in the number 
of RGB stars. This evidence is supported by the substantial change 
in the slope of the cumulative LF plotted in the bottom panel of Figure 2,
and can also be identified in the CMDs plotted in Figure 1. 
To assess the nature of this feature, Figure 2 shows the comparison 
between the observed differential LF and the theoretical LF for an age 
of 14 Gyr. We find that for magnitudes fainter than $I\approx22.4$ mag, 
the observed stellar counts are, at the $2\sigma$ level, larger than 
the theoretical counts expected for an old stellar population.

This finding could be interpreted as the evidence of a secondary sample 
of RGB stars connected with a stellar population younger than the main 
RGB stellar component. 
In fact, for stellar ages lower than $\approx1Gyr$, the TRGB luminosity 
becomes quite sensitive to the age.
To constrain the age of such a population we take into account the 
I magnitude difference between the RGB tip associated with the oldest 
component, located at $I_{TRGB}=21.65$ mag, and the LF discontinuity 
located at fainter magnitudes i.e. $I\approx22.4$mag. By adopting this 
approach and by assuming for the younger stellar component, the same
metallicity of the oldest one, we estimate that its age is $\approx0.7$ Gyr.

%%%%%%%%%%%%%%%%%%%%%%%%%%%%%%%%%%%%%%%%%%%%%%%%%%%%%%%%%%%%%%%%%%%%%%%%%%%
\section{Some hints on the Antlia-Sextans grouping.}

One of the main reason why the global properties of Antlia are so interesting  
is because it should be located beyond the zero-velocity surface of the LG, 
and therefore its motion can be used for providing independent estimates of
both the LG dark matter halo and the age of the Universe (Lynden-Bell 1981).  
However, the location of this galaxy within the LG is still controversial.
In fact, SCO suggested on the basis of its position in the heliocentric 
radial velocity versus apex angle diagram that it is located near the 
outer edge of the LG. At the same time, it was pointed out by ADGMD 
on the basis of the relative velocity between Antlia and NGC3109 that 
it is unlikely that this pair of galaxies is gravitationally bound. 
On the other hand, Yahil, Tammann, \& Sandage (1977), 
Lynden-Bell \& Lin (1977) and, more recently, VDB in a detailed 
analysis of the nearest group of galaxies 
brought out that Antlia together with Sextans A/B and NGC3109 forms a 
small cluster of galaxies which is not bounded to the LG, and therefore 
that it is expanding with the Hubble flow. He also suggested that
Antlia is probably a satellite of NGC3109 and that this pair to 
be gravitationally stable should contain a sizable amount of dark 
matter.  

By adopting the distance moduli with the relative errors estimated 
by SC98 (see their Table 2, column 9 and Table 3, column 2) 
by means of the TRGB method for the other three members of 
this group we find that the corresponding distances are:  
D(Sextans A)=$1.51\pm0.1$ Mpc, D(Sextans B)=$1.45\pm0.09$ Mpc, 
and D(NGC3109)=$1.37\pm0.09$ Mpc. Taken at face values these 
distances together with the Antlia distance suggest that within 
the errors these four galaxies are probably located at the same 
distance, thus supporting the finding by VDB that they form 
a small nearby clustering. Note that with the exception of NGC3109 these 
distance estimates are systematically larger than those adopted 
by VDB. The discrepancy ranges from 5\% for Sextans A up 
to 13\% for Antlia. The reason for the disagreement is partially 
due to the difference in the TRGB calibration and partially to the 
different standard candles adopted by VDB 
(classical Cepheids and TRGB method). The main advantage of 
our distance determinations is that they are based on the same 
standard candle and on the same TRGB calibration. 
However, distance determinations based both on the LFM and on 
the SC98 calibration agree within their errors with the distance 
scale based on Cepheid Period-Luminosity (PL) relation. 
This problem might be resolved by a detailed comparison of 
distance determinations based on Cepheid PL relations which 
account for the metallicity dependence and on TRGB method 
(Bono, Marconi, \& Stellingwerf 1999). 

In order to disentangle this thorny problem we estimated the distance 
of NGC3109 by adopting the sample of classical Cepheids observed
in this galaxy by Musella, Piotto, \& Capaccioli (1998) and the 
theoretical $PL_I$ and $PL_V$ relations for Z=0.004 provided by 
Bono et al. (1999). Interestingly enough, we find that the reddening
corrected distance moduli in these two bands are $25.8\pm0.1$ mag 
and $25.82\pm0.08$ mag respectively. 
Within the errors, which account only for the intrinsic
dispersion, these distance determinations seem to support more the 
SC98 than the LFM calibration of the TRGB method. In fact, SC98 derived 
a distance modulus for NGC3109 of $25.69\pm0.14$ mag which is in good 
agreement with the Cepheid distance, while Lee (1993) by adopting the 
LFM calibration found a distance modulus of $25.45\pm0.15$ mag 
and the corresponding distance is 15\% smaller than the Cepheid 
distance. 
On the basis of this finding, it goes without saying that DGs 
which host both young and old stellar populations can play a key role to 
settle down the dependence of Cepheid distance scale on metallicity 
since these stellar systems are characterized by a much smaller 
metallicity gradient when compared with large spiral galaxies 
(Mateo 1998). 

By taking into account the TRGB distances of NGC3109 and 
Antlia and their separation on the sky ($\approx 1^\circ.18$) we 
estimated that the projected distance between these galaxies 
is $\approx 140$ kpc. The projected distance decreases to 
$\approx 70$ kpc if we use the new NGC3109 distance based on 
Cepheid distance. 
By adopting these two different projected distances and the equation 
(4) given by VDB we find that this system should have a total mass 
larger than $7.8\times10^{10} M_\odot$ (TRGB distances) and 
$4.0\times10^{10} M_\odot$ (Cepheid distance to NGC3109) to be bound.
These values translates, by assuming $m_{V,0} (Antlia)=15.58$ (ADGMD) and
$m_{V,0} (NGC3109)=9.63$ (Carignan 1985; Minniti et al. 1999), into 
total mass-to-light ratios of $(M/L_V)_0 \ge 350$ and 170 in solar 
units. Taken at face value these ratios imply that this system should 
contain an amount of dark matter which is at least a factor of 2-4 
larger than in any other dwarf in the LG (see Table 4 in Mateo 1998). 
Therefore, it is seems unlikely that these two galaxies are 
gravitationally bound. Finally, we mention that the increase in the 
mean distances supplies a straightforward support to the evidence 
brought out by VDB that the Ant-Sex clustering is located beyond the 
zero-velocity surface of the LG.

%%%%%%%%%%%%%%%%%%%%%%%%%%%%%%%%%%%%%%%%%%%%%%%%%%%%%%%%%%%%%%%%%%%%%%%%%%
\section{Summary and conclusions.}
 
Two CMDs -$(I, V-I)$ and $(I, B-I)$- together with the LF in the I band, 
based on photometric data collected with FORS I during 
the SV program of the VLT were used for constraining the global 
properties of Antlia. The new data confirm, as originally suggested 
by SCO and ADGMD, that low-mass, metal-poor stars with an age of the
order of 10 Gyr are the main stellar component of this galaxy and that 
young blue stars are located close to the center. The presence of 
interstellar dust suggested by SCO is not confirmed by current 
photometric data. 
 
The comparison between theory and observations suggests that the 
young stellar component is characterized by an age ranging from 
100 to 150 Myr, and in turn that in this galaxy should be present 
classical Cepheids with periods of the order of few days. 
By adopting the calibrations of the TRGB method and of the color 
index $V-I$ vs. metallicity suggested by SC98 we estimated that 
the distance modulus of Antlia is $(m-M)_0=25.89\pm0.10$ mag 
-i.e. $D=1.51\pm0.07$ Mpc-, while its mean metallicity is 
$[M/H]=-1.3\pm0.15$. This distance estimate is 13\% larger than 
the distance determinations provided by SCO and ADGMD, while the 
mean metallicity is 0.3 dex more metal-rich than the value 
suggested by ADGMD. The disagreement with previous estimates 
available in the literature is mainly due to systematic differences
in the calibration of the TRGB method and of the color index vs. 
metallicity relation provided by LFM and SC98.   

Interestingly enough, we find that the differential LF shows a 
secondary peak at $I\approx22.5$ mag which is at $2\sigma$ level 
larger than theoretical predictions. We suggest that this feature 
could be due to a secondary young/intermediate-age stellar component. 
By assuming that this sample of RGB stars presents the same metallicity 
of the old component we estimated that its age should be $\approx0.7$ Gyr.    
This evidence together with the appearance of the blue stars suggests 
that after the initial burst who took place approximately  10 Gyr ago 
this galaxy experienced two further star formation episodes 
$\approx0.7$ and $\approx0.1$ Gyr ago.   
 
Finally by using the TRGB method we derived in a homogeneous context 
the distances of Sextans A/B, and NGC3109 we find that these three
galaxies together with Antlia are located within the errors at the 
same distance. Thus supporting the finding by VDB  that these 
galaxies form a small nearby clustering. The new distances also 
support the evidence that this grouping could be located beyond 
the zero-velocity surface of the LG (VDB).  
Obviously new observations aimed at detecting both horizontal branch 
stars and RR Lyrae stars as well as at detecting and measuring 
classical Cepheids can supply fundamental constraints on the intrinsic 
distance and on the global properties of this intriguing neighborhood.

\begin{acknowledgements}
It is a pleasure to thank V. Castellani and M. Marconi for many 
interesting discussions on an early draft of this paper. We also 
warmly acknowledge M. Salaris for many stimulating suggestions 
on the content of the paper. Detailed and pertinent comments 
from the referee, Sydney van den Bergh, have contributed to 
improving the content of this paper.  
\end{acknowledgements}

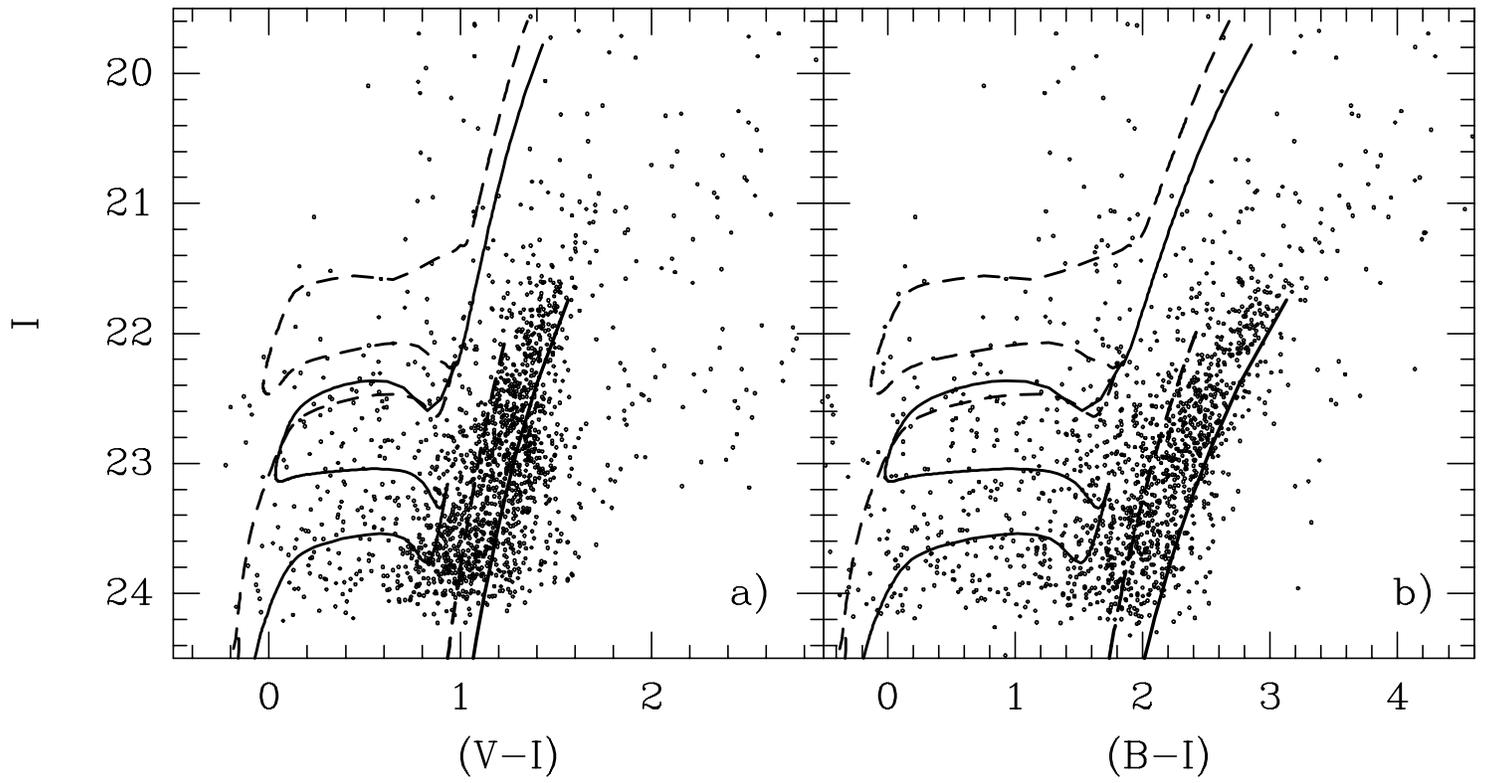
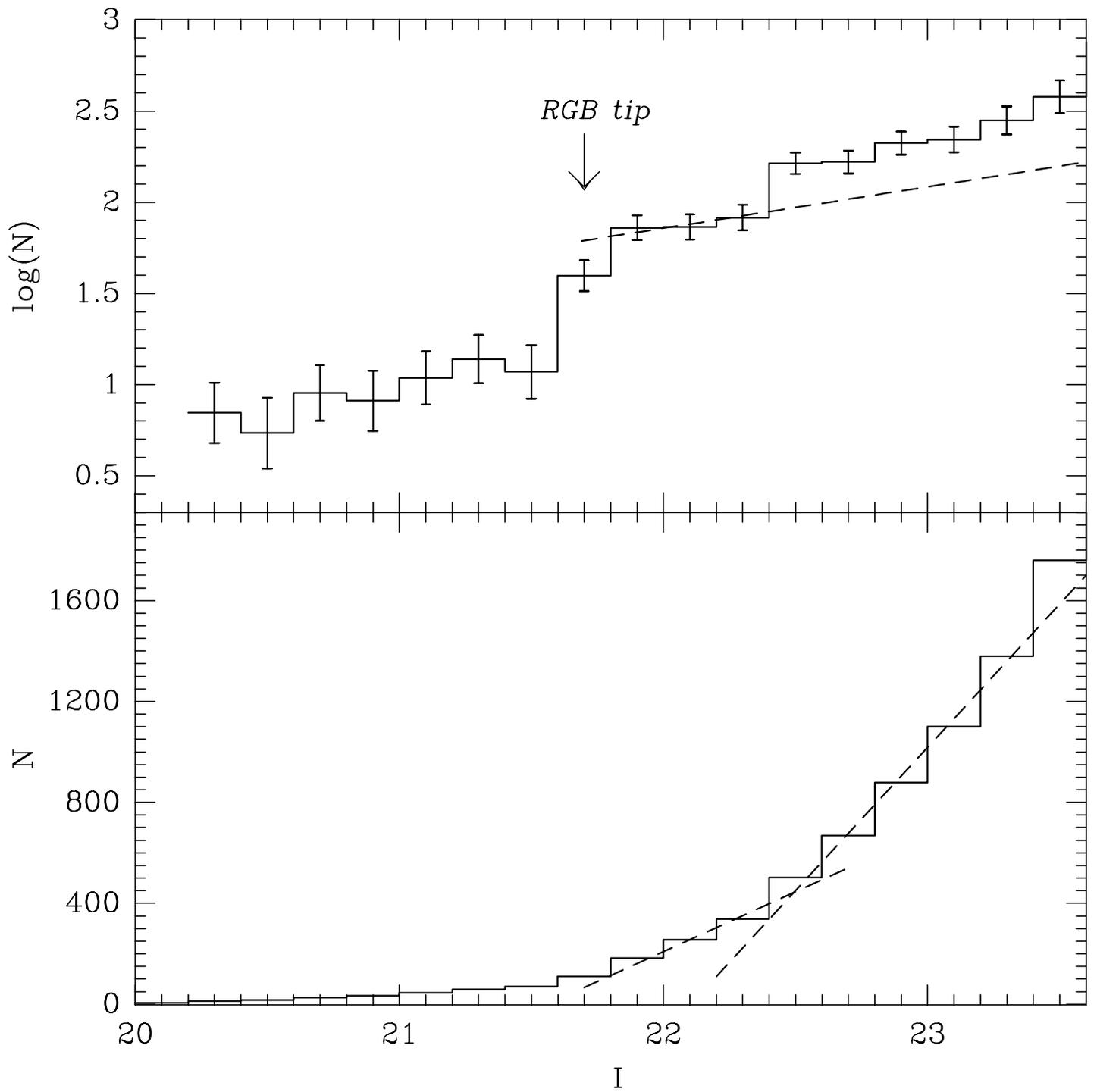
\end{document}